\documentclass[12pt]{article}
\usepackage{latexsym,amsfonts,amsmath,amsthm,amssymb}

\newcommand{\la}{\lambda}
\newcommand{\de}{\delta}

\newcommand{\Om}{\Omega}
\newcommand{\om}{\omega}
\newcommand{\F}{\cal {F}}
\newcommand{\A}{\cal {A}}
\newcommand{{\bP}}{\bf {P}}
\newcommand {\PP}{{\bP}(b=b_i/a=a_n, C)}
\newcommand {\Pp}{{\bP}(b=b_i/a=a_n)}

\title{Contextual approach to quantum mechanics and the theory of the fundamental prespace}
\author{Andrei Khrennikov\footnote{International Center for Mathematical Modeling 
in Physics and Cognitive Sciences, Andrei.Khrennikov@msi.vxu.se; supported by EU-Network
 "QP and Applications} \\
MSI, University of V\"axj\"o, S-35195, Sweden}
\begin{document}
\maketitle

\begin{abstract}We constructed a Hilbert space representation of 
a contextual Kolmogorov model. This representation is based on two fundamental
observables -- in the standard quantum model these are the position and momentum observables.
This representation has all distinguishing
features of the quantum model. Thus in spite all ``No-Go'' theorems (e.g., 
von Neumann, Kochen and Specker,..., Bell) we found the realist basis of 
quantum mechanics. Our representation is not standard model with hidden variables.
In particular, this is not a reduction of the quantum model to the classical one.
Moreover, we see that such a reduction is even in principle impossible. 
This impossibility is not the consequence of a mathematical theorem but it follows
from the physical structure of the model. By our model quantum states are very
rough images of domains in the space of fundamental parameters - PRESPACE. Those
domains represent complexes of physical conditions.
By our model both classical and quantum physics describe REDUCTION
of PRESPACE-INFORMATION.  Quantum mechanics is not complete. In particular, there are 
prespace contexts which can be represented only by a so called hyperbolic quantum model.
We predict violations of the Heisenberg's uncertainty principle and existence
of dispersion free states.
\end{abstract}
 
 \section{Introduction}

Since the early days of quantum mechanics (QM) there have been permanent discussions on the problem: 

{\bf What is really QM about? }

In particular, the question of the greatest importance is 

{\bf Can QM be interpreted in a realist way?}

We recall that A. Einstein was sure that a realist interpretation of QM could be found:
QM is not complete and there would be found finer, 
``prequantum", descriptions of physical reality. The most well known consideration 
of these problems was presented in the famous EPR paper [1]. This paper induced intensive discussions 
and, finally, various ``no-go" theorems. The ``final no-go theorem" of J. Bell [2] induced a rather common 
opinion that QM cannot be based on a {\it local realist model.}
Personally I do not think that locality (in the physical space ${\bf R}^3)$ really
plays the fundamental role in the realist interpretation of QM\footnote{Since this paper is not about EPR, 
I would no like to go deeply into detail. It should be remarked that it is impossible
even to discuss locality in Bell's framework, because space variables 
do not present in Bell's model of ``local reality''
and Bell's inequality says nothing about those
variables, see [3] for detail. Regarding the original EPR experiment I remark that a 
local realist model was recently constructed, see [3].}. Therefore I would like to come back to 
the original attempts of Einstein to find violations of Heisenberg's uncertainty relation. 
This was the crucial point of investigations of the possibility of a realist interpretation of QM. 
Unfortunately (as it was already mentioned) the main stream of those investigations was later directed to
the EPR-Bohm-Bell framework.

We underline that the possibility to violate Heisenberg's uncertainty principle for
two fundamental variables, e.g., position and momentum, need not imply the realization of the program 
of the hidden variables (HV) reconstruction of QM. The latter can be formulated (see J. Bell [4]):

{\small ``The question at issue is whether the quantum mechanical states can be regarded as ENSEMBLES of
states further specified by additional variables, such that given values of these variables together
with the state vector determine precisely the results of individual measurements."} 

Those hypothetical well-specified states are said to be {\it `dispersion free'.} In this paper we present a model 
in which dispersion free states exist for two fundamental variables $b$ and $a$ (represented by
noncommutative operators $\widehat b$ and $\widehat a$) but for an arbitrary quantum observable
such states need not exist. Of course, an orthodox adherent of the HV-approach would not be so much
interested in our model. But we could not do anything more. This is the reality.

Moreover, even the general ideology of HV reduction of QM has not so much to do with our approach. 
We are not looking for some HV-models for QM. In the HV-approach the main problem is not nonexistence 
of HV-models, but existence of too many models. Many of them are totally meaningless from the physical
viewpoint, see, e.g., Bell's example [1] for the two dimensional HV-system. 
The HV-reductionist approach has also induced numerous  
discussions which kinds of reductions are acceptable -- 
von Neumann's ``no--go''  and 
Kochen and Specker's  theorems (see [5], [6], and see [7]-[9] for reviews), De Broglie double solution
model [10], Bohmian mechanics [11], [12], and Nelson-Guerra stochastic mechanics, see, e.g., [13].

I think that the starting point should be not QM. The QM-formalism by itself could not tell so much about 
features of a prequantum realist model which   is (roughly) encoded by this formalism. Inversely we should start with 
a realist model and try to find a ``natural representation" of such a model in a Hilbert space. Since QM is 
a statistical theory such a natural representation should be a probabilistic one. Roughly speaking the problem 
is to find a natural way to represent Kolmogorovian probabilities by complex amplitudes (or in the abstract 
framework by normalized vectors in a Hilbert space). In a series of papers [14]---[17] there was shown that 
such a representation can be constructed on the basis of a so called 
{\bf{contextual formula of total probability}} for observables $b$ and $a:$
\begin{equation}
\label{c1}
p_C^b(x)=\sum_y \; p_C^a(y) \;p^{b/a}(x/y)\; + 
\end{equation}
$$
2\sum_{y_1 < y_2}\sqrt{p_C^a(y_1)\;p_C^a(y_2)\;p^{b/a}(x/y_1)\;p^{b/a}(x/y_2)}\; \cos \theta^{(y_1y_2)}_C(x),
$$
where $p_C^a(y)={\bf P}(a=y/C), p_C^b(x)={\bf P}(b=x/C)$ are probabilities to observe 
values $a=y$ and $b=x$ under the complex of physical conditions -- {\bf context} -- 
$C$ and $p^{b/a}(x/y)={\bf{P}}(b=x/a=y)$ are transition probabilities. A complex amplitude 
$\varphi_C^{b/a}(x)$ corresponding to the representation (\ref{c1}) gives the QM-representation 
of context $C.$ In [17] it was shown that for a Kolmogorov probability space 
$
{\cal K}=(\Omega, {\cal F}, {\bf P})
$
and a pair of {\it incompatible Kolmogorovian random variables} $b$ and $a$ we can construct a
natural quantum representation. This representation is rigidly based on a pair of
variables $b$ and $a$ --- fundamental  (for that
concrete representation of physical reality) observables. In particular, the standard quantum 
representation is based on the {\bf position and momentum observables.}
There exists a map $J^{b/a}$ which maps contexts (represented by subsets of $\Omega$) 
into quantum states $\equiv$ complex $b/a$-transition amplitudes of 
probability\footnote{In some sense we came back to the original Hilbert's viewpoint to a
wave function as a transition amplitude, see [18], see also Lande [19], cf. Gudder [20], Accardi [21], 
Ballentine [22], 
Khrennikov [14]-[17].}. 

Points of $\Omega$ are interpreted as {\it fundamental physical parameters}\footnote{If you like HV... 
But the general HV-approach was so discredited  by former investigations (since people wanted too much 
for such a HV-description) that we would not like even to refer to HV.}. 
We call $\Omega$ {\bf prespace} and fundamental parameters --- prepoints.

The main distinguishing feature of the representation map $J^{b/a}$ is the huge 
{\bf compression of information.} In particular, every point represented in the conventional 
mathematical model of physical space by a vector $x \in {\bf R}^3$  
is the image of a subset 
$$
B_x=\{\omega \in \Omega:b(\omega)=x\}
$$ 
of $\Omega$ which can contain millions of prepoints. In the conventional quantum representation
of the prespace the fundamental variable $b=q$ is the position observable. We have a similar picture 
for the momentum observable. In the quantum model we consider ``classical physical points" $x\in {\bf R}^3$ 
as represented by eigenvectors of the position operator. Thus by going from the 
``classical physical space" ${\bf R}^3$ to the quantum physical (Hilbert) space $H$ and then to the prespace 
$\Omega$ we obtain finer and finer descriptions of reality. 

Another distinguishing feature of 
the $J^{b/a}$-representation of the prespace $\Omega$ in the Hilbert space $H$ is the creation 
of superpositions of ``classical states". The origin of the quantum superposition can be very easily 
explained by our prespace model. For example, let us consider a context $C \subset \Omega$ such 
that $C \subset B_{x_1} \cup B_{x_2}, x_1, x_2 \in {\bf R}^3, x_1 \ne x_2,$ 
but neither  $C \subset B_{x_1}$ nor $ C \subset B_{x_2.}$ 
The image $\varphi_C=J^{b/a}(C) \in H$ is a quantum state describing a quantum system which is 
``in a superposition of the positions'' $x_1$ and $x_2.$ 

Thus inspite of all ``no-go" theorems (e.g., 
von Neumann, Kochen and Specker, Bell, \ldots), we have constructed a realist model of QM. 
In this model (as it was wanted by A. Einstein) {\it the Heisenberg uncertainty relation can be violated} 
for fundamental observables (e.g., the position and momentum) which are used for our classical and
quantum representations of reality. Points ({$\omega \in \Omega$}) of the prespace are 
{\bf dispersion free states.}

In our model only the fundamental observables correspond to random variables on the prespace.
Other quantum and classical observables have only some indirect relation to random variables on the prespace. 
So we could not consider such, e.g., quantum observables as real observables -- functions of fundamental
parameters.
Nevertheless, for a wide class of quantum observables (including QM-Hamiltonians) we 
have the coincidence of averages with averages of corresponding random variables on $\Omega$. 
Here we speak about averages with respect to the state $\varphi_C=J^{b/a}(C)$ and context $C,$ 
respectively. In our model only quantum observables belonging to a special class 
(class ${\cal O}_+(a,b))$ have realist interpretation.

We underline that our investigations has nothing to do with attempts to find some general probabilistic
model which would contain Kolmogorov as well as quantum probabilities as particular cases, cf., e.g.,
Mackey [24], Gudder [20], Ludwig [25], Devies and Lewis [26], Accardi [21], 
Ballentine [22] ...,  Khrennikov [27], Hardy [28]. For us the main distinguishing 
feature of quantum theory is not 
a new (``quantum'') behaviour of probabilities, but a special way of representation of (ordinary)
probability. 

By our model dispersion free states (for, e.g., position and momentum observables)
can exist and {\bf the Heisenberg's uncertainty principle can be violated.}

\section{Contextual formula of total probability}

Let $(\Om, \F, {\bP})$ be a Kolmogorov probability space, [29].

By the standard Kolmogorov axiomatics sets $A\in {\cal F}$ represent {\it{events.}} 
 In our simplest model of {\it{contextual probability}} (Kolmogorovian contextual model)
 the same system of sets, ${\cal F}$, 
 is used to represent complexes of experimental physical conditions -- {\it{contexts.}}
 We can consider a set  $C \in {\cal F}$ as a collection of physical parameters $\omega$ describing 
 a complex of physical conditions. This is a context--interpretation of sets $C \in {\cal F}.$
 
 By the event--interpretation of sets $A \in {\cal F}$ such a set $A$ is a collection of 
 physical parameters inducing the corresponding event (denoted by the same symbol $A$).
 
 In principle, in a mathematical model events and contexts can be represented 
 by different families of sets, see, e.g., Renye's model. 
 We will not do this from the beginning. But later we will fix a proper subfamily 
 of contexts ${\cal C} \subset {\cal F}$. 
 
 The conditional probability is mathematically
 defined by the Bayes' formula:
 \[{\bf P}(A/C)=\frac{{\bf P}(AC)}{{\bf P}(C)}, {\bf P}(C) \ne 0.\]
In our contextual model this probability has the meaning of the probability of occurrence of the event
$A$ under the complex of physical conditions $C.$ Thus it is not the probability of occurrence
of the event $A$ under the condition that the event $C$ has occurred (as it is 
assumed in the Kolmogorov theory). \footnote{The reader might think that the difference in interpretations
is negligible. But I would like to underline that this is the crucial point of all our considerations.} 

Thus it would be more natural to call ${\bf P}(A/C)$ a {\it{contextual probability}}
and not {\it conditional probability.} Roughly speaking to find ${\bf P}(A/C)$ we should
find parameters $\omega^A$ favouring to the occurrence of the event $A$ among parameters
$\omega^C$ describing the complex of physical conditions $C.$

Let ${\A}=\{ A_n \}$ be finite or countable complete group of inconsistent contexts: 
$$
A_i A_j= \emptyset, i \not= j,\;\;\;\;  \cup_i A_i=\Om.$$ 
Let $B\in {\cal F}$  be an event and $C\in {\cal F}$ be a context and let ${\bP}(C)>0.$ 
We have the standard formula of total probability, see, e.g., [17]:
$$
{\bP}(B/C)=\frac{{\bP}(BC)}{{\bP}(C)}=
\sum_n \frac{{\bP}(BA_nC) {\bP}(A_nC)}{{\bP}(C) {\bP}(A_nC)}
$$
(if ${\bP}(A_nC)>0$ for all $n)$  and hence
\begin{equation}
\label{CONV}
{\bP}(B/C)=\sum_n {\bP}(A_n/C){\bP}(B/A_nC)
\end{equation}
Of course, in the conventional Kolmogorov model we operate only with events. Thus inspite of using
the standard Kolmogorov measure-theoretical probabilistic formalism, from the very beginning we use
a new interpretation of conditioning in this formalism. Instead of the conventional even-conditioning,
we use {\bf context-conditioning.} Thus there is nothing new from the mathematical viewpoint and the
reader may be curious: Is it possible to find something new by using the same mathematical apparatus
and by changing only the interpretation? Yes, we shall construct totally new representation of the 
Kolmogorov model in a Hilbert space. This representation is nontrivial -- Kolmogorovian (but contextual)
random variables are represented by in general noncommutative operators.

In particular, let $a$ and $b$ be discrete random variables taking values 
$a_i, i=1, \ldots , k_a$ and $b_j, j=1, \ldots, k_b,$ where $k_a, k_b < \infty.$ We have 
$$
{\bP}(b=b_i/C)=\sum_n {\bP}(a=a_n/C) {{\bP}(b=b_i/a=a_n, C)} \;.
$$

Let a measurement of the variable $a$ disturb essentially physical systems 
$\om \in  \Om.$  Let us fix some complex of conditions ({\it context}) $C,$ 
see [14]--[17] for detail. One cannot  measure $b$ and $a$ simultaneously in the context
$C.$ Thus the probabilities 
${\bP}(b=b_i/a=a_n, C)$ are ``hidden" (or ontic) probabilities. \footnote{We  are not able
to select parameters $\omega^{A_n}$ favouring to the realization of $a=a_n$ 
without to disturb context $C.$}
However, we can measure the variable $b$ in the context
$A_n= \{\omega : a(\omega) =a_n\}.$ Thus we can not prepare for the context $C$ systems $\om$ such that we know 
that simultaneously $b(\om)=b_i, a(\om)=a_n,$ but we can prepare systems 
$\om$ such that $a(\om)=a_n$ and in this context we can perform the 
$b-$measurement. Hence the probabilities ${\bP}(b=b_i/a=a_n)={{\bP}(B_i/A_n)}$ are 
well defined. Here 
$$
B_i=\{\om \in \Om: b(\om)=b_i\}\; \;  \mbox{and} \; \;
A_n=\{\om \in \Om: a(\om)=a_n\} .
$$
I would like to modify the formula of total probability (\ref{CONV}) by 
eliminating hidden probabilities $\PP$ and using only observable 
probabilities $\Pp$.

{\bf Definition 1.} (Context) {\it{A set $C$ belonging to $\F$ is said to 
be a context with respect to  a complete group of inconsistent contexts ${\A}=\{A_n\}$ if 
${\bP}(A_nC)\not =0$ for all $n.$}}

We denote the set of all ${\A}-$contexts by the symbol  ${\cal C}_{\A}.$

{\bf Definition 2.} {\it{Let ${\A}=\{A_n\}$ and ${\cal B}=\{B_n \}$ be two 
complete groups of inconsistent contexts. They are said to be incompatible if 
${\bP}(B_n A_k) \not = 0$ for all $n$ and $k.$}}

Thus ${\cal B}$ and ${\cal A}$ are incompatible iff every $B_n$ is a context with respect to 
${\cal A}$ and vice versa, see Appendix 1 for detail.

Random variables $a$ and 
$b$ inducing incompatible complete groups ${\cal A}=\{A_n\}$ and ${\cal 
B}=\{B_k\}$ of inconsistent contexts are said to be {\it {incompatible random 
variables.}}

{\bf Theorem 1.} (Interference formula of total probability) 
 {\it Let $\A$ and ${\cal B}$  be incompatible and let $C$ be a context with respect to $\A.$
 Then the following ``interference formula of 
total probability'' holds true for any $B\in {\cal B}:$}
\begin{equation}
\label{INN}
{\bP}(B/C)=\sum {\bP}(A_n/C){\bP}(B/A_n)+
\end{equation}
\[2\sum_{n<m}\lambda_{nm}(B/{\A},C)\sqrt{{\bP}(A_n/C){{\bP}(A_m/C) 
{\bP}(B/A_n){\bP}(B/A_m)}}\]
where
$$
\lambda_{nm}(B/{\cal A};C)=
\frac{\delta_{nm}(B/{\cal A};C)}{2\sqrt{{\bP}(A_n/C){\bP}(B/A_n){\bP}(A_m/C){\bP}(B/A_m)}}
$$
and 

$
\delta_{nm}(B/{\cal A};C)
$
\begin{equation}
\label{D1}
=\frac{[{\bP}(A_n/C)({\bP}(B/A_nC)-{\bP}(B/A_n))+{\bP}(A_m/C) ({\bP}(B/A_mC)-  {\bP}(B/A_m)]}{k_a-1}
\end{equation}

{\bf Proof.} We have:
$$
{\bP}(B/C)=\sum_n{\bP}(A_n/C)({\bP}(B/A_nC)+{\bP}(B/A_n)-{\bP}(B/A_n))
$$
$$
= \sum_n{\bP}(A_n/C){\bP}(B/A_n)+\delta(B/{\cal A},C),
$$
where
\begin{equation}
\label{SD}
\delta(B/{\A},C)= \sum_n{\bP}(A_n/C)({\bP}(B/A_nC)-{\bP}(B/A_n)).
\end{equation}
Finally, we remark that we can represent the perturbation term as the sum of perturbation
terms corresponding to pairs of $(A_n, A_m):$
$$
\delta(B/{\A},C)=\sum_{n<m} \delta_{nm}(B/{\cal A};C),
$$ 
where $\delta_{nm}(B/{\cal A};C)$ is given by (\ref{D1}).

The $\la_{nm}(B/{\A},C)$ are called  {\it coefficients of 
statistical disturbance.} Coefficients $\la_{nm}(B/{\A},C)$ describe disturbances of 
probabilities induced by filtrations with respect to values $a=a_n$ in the 
context $C.$ Depending on magnitudes of these coefficients we can rewrite 
the nonconventional formula of total probability in various forms that are 
useful for representing (\ref{INN}) as a transformation in a complex linear 
space or a Clifford modular, see [14]-[17] for the details.

In our further investigations we will use the following result:

{\bf Lemma 1.} {\it{Let conditions of Corollary 1. hold true. Then}}
\begin{equation}
\label{CD1}
\sum_k \de(B_k/{\A},C)=0
\end{equation}

{\bf Proof.} We have $1=\sum_k {\bP}(B_k/C)=\sum_k \sum_n {\bP}(A_n/C)
{\bP}(B_k/A_n)+\sum_k \de(B_k/{\A},C).$
But $\sum_n(\sum_k {\bP}(B_k/A_n)) {\bP}(A_n/C)=1.$

As a consequence of this lemma we have:
\begin{equation}
\label{CD2}
\sum_k \sum_{l<m}\la_{l m}(B_k/{\A},C)\sqrt{{\bP}(A_l/C){\bP}(A_m/C) 
{\bP}(B_k/A_l) {\bP}(B_k/A_m)}=0
\end{equation}

1). Suppose that $a=a_n$ filtrations (in the context $C$)\footnote{First we prepare a statistical ensemble $O_C$
of physical systems $\omega$ under the complex of (e.g., physical) conditions $C.$  Then we perform
a measurement of the random variable $a$ for elements of the ensemble $O_C.$ Finally, we select all systems
for which we obtained the value $a=a_n.$}
induce statistical 
disturbances having relatively small coefficients $\la_{nm}(B/{\A},C),$ 
namely, for every $B\in {\cal B}$
\[\vert \la_{nm}(B/{\A},C)\vert\leq 1 \;.\]

In this case we can introduce new statistical parameters $\theta_{nm}(B/{\A},C)\in 
[0,\pi]$ and represent the coefficients of statistical disturbance in the 
trigonometric form: 
$$
\la_{nm}(B/{\A},C)=\cos \theta_{nm}(B/{\A},C).
$$
Parameters $\theta_{nm}(B/{\A},C)$ are said to be {\it{relative phases}} of 
an event $B$ with respect to a complete group of inconsistent events ${\A}$ 
(in the context $C$).

In this case we obtain the following interference formula of total probability:
$$
{\bP}(B/C)=\sum_n {\bP}(A_n/C) {\bP}(B/A_n) 
$$
\begin{equation}
\label{TNC}
+ 2\sum_{n<m} \cos \theta_{nm}(B/{\A},C) (B/{\A},C)
\sqrt{{\bP}(A_n/C) {\bP}(A_m/C) {\bP}(B/A_n) {\bP}(B/A_m).}
\end{equation}
This is nothing other than the famous {\it formula of interference of 
probabilities.}\footnote{Typically this formula is derived by using the Hilbert 
space (unitary) transformation corresponding to the transition from
one orthonormal basis to another and Born's probability postulate.
The orthonormal basis under quantum consideration consist of eigenvectors of
operators (noncommutative) corresponding to quantum physical observables 
$a$ and $b.$} We demonstrated that in the opposite of the common (especially in quantum 
physics) opinion nontrivial interference of probabilities need not be 
related to some non-Kolmogorovian features of a 
probabilistic model. In our considerations everything is Kolmogorovian. 
Interference of probabilities is a consequence of the 
impossibility of using conditioning with respect to $\{a=a_n, C\}$ 
(to combine two contexts -- $C$ and $a)$ for 
random variables $a$ which measurement disturbs essentially physical 
systems $\om \in \Om.$ 

Starting from (\ref{TNC}) we shall derive (for dichotomous random variables)
Born's rule, construct for any context $C$ a complex probability amplitude,
introduce a Hilbert space structure on the space of complex amplitudes and 
represent random variables on the Kolmogorov probability space by (in general 
noncommutative) operators in the Hilbert space.

2). Suppose that $a=a_n$ filtrations 
induce statistical 
disturbances having relatively large coefficients $\la_{nm}(B/{\A},C),$ 
namely, for every $B\in {\cal B}$
\[\vert \la_{nm}(B/{\A},C)\vert\geq  1 \;.\]

In this case we can introduce new statistical parameters $\theta_{nm}(B/{\A},C)\in 
[0,+ \infty]$ and represent the coefficients of statistical disturbance in the 
trigonometric form: 
$$
\la_{nm}(B/{\A},C)=\pm \cosh \theta_{nm}(B/{\A},C).
$$
Parameters $\theta_{nm}(B/{\A},C)$ are said to be hyperbolic {\it{relative phases}} of 
an event $B$ with respect to a complete group of inconsistent events ${\A}$ 
(in the context $C$).

In this case we obtain the following interference formula of total probability:
$$
{\bP}(B/C)=\sum_n {\bP}(A_n/C) {\bP}(B/A_n) 
$$
\begin{equation}
\label{TNC1}
\pm 2\sum_{n<m} \cosh \theta_{nm}(B/{\A},C) (B/{\A},C)
\sqrt{{\bP}(A_n/C) {\bP}(A_m/C) {\bP}(B/A_n) {\bP}(B/A_m).}
\end{equation}

3). Suppose that $a=a_n$ filtrations 
induce for some $n$  statistical 
disturbances having relatively small coefficients $\la_{nm}(B/{\A},C)$
and for other $n$  statistical 
disturbances having relatively large coefficients $\la_{nm}(B/{\A},C).$ 
Here we have the interference formula of total probability containing 
trigonometric as well as hyperbolic interference terms.

\section{Dichotomous random variables.}

We study only models with {\bf trigonometric interference.}
We set 
$$
{\cal C}=\{ C\in {\cal C}_{{\cal A}}: \vert\lambda(B_j/{\cal A}, C)\vert\leq 1\}
$$
We call elements of ${\cal C}$ trigonometric contexts. We shall see that QM can be interpreted
as a representation of trigonometric contexts. We can also introduce hyperbolic contexts
which can be represented in a hyperbolic Hilbert space, see [30].

{\bf 3.1. Interference and complex probability amplitude, Born's rule.}
Let us study in more detail the case of incompatible dichotomous random 
variables $a=a_1, a_2, b=b_1, b_2.$ We set $Y=\{a_1, a_2\}, X=\{b_1, 
b_2\}$ (``spectra'' of random variables $a$ and $b).$ 
Let $C\in {\cal C}$ be a context for both random variables $a$ and $b.$ We set
\[p_C^a(y)={\bP}(a=y/C), p_C^b(x)={\bP}(b=x/C), p(x/y)={\bP}(b=x/a=y),\]
$x \in X, y \in Y.$
The interference formula of total probability (\ref{TNC}) can be written in 
the following form
\begin{equation}
\label{Two}
p_c^b(x)=\sum_{y \in Y}p_C^a(y) p(x/y) + 2\cos \theta_C(x)\sqrt{\Pi_{y \in 
Y}p_C^a(y) p(x/y)}\;,
\end{equation}
where $\theta_C(x)=\theta(b= x/{\cal A}, C)= \arccos \lambda(b=x/{\cal A}, C), x \in X, C\in {\cal C}.$
We remark that in the case of dichotomous random variables: 
$$
\delta(b=x/{\cal A}, C)=p_c^b(x)-\sum_{y \in Y} p_C^a(y)p(x/y)
$$ 
and 
$$
\lambda(b=x/{\cal A}, C)  =\frac{\delta(b=x/{\A}, C)}{2\sqrt{\Pi_{y\in Y}p_C^a(y)p(x/y)}} .
$$
By using the elementary formula: 
$$
D=A+B+2\sqrt{AB}\cos \theta=\vert \sqrt{A}+e^{i\epsilon \theta}\sqrt{B}|^2, 
$$ 
for $A, B > 0, \epsilon=\pm 1, \theta\in [0,\pi].$
we can represent the probability $p_C^b(x)$ as the square of the complex amplitude:
\begin{equation}
\label{Born}
p_C^b(x)=\vert\varphi_C(x)\vert^2 \;.
\end{equation}
We fix some pair of signs $\epsilon(x), x\in X$ (e.g., $\epsilon(b_1)=-1$ and  $\epsilon(b_2)= +1 \;).$
We set
\begin{equation}
\label{EX1}
\varphi(x)\equiv \varphi_C(x)=\sqrt{p_C^a(a_1)p(x/a_1)} + e^{\epsilon(x) \theta_C(x)} \sqrt{p_C^a(a_2)p(x/a_2)} \;.
\end{equation}
We denote the space of functions: $\varphi: X\to {\bf C}$ by the symbol 
$E=\Phi(X, {\bf C}).$ Since $X= \{b_1, b_2 \},$ the $E$ is the two dimensional 
complex linear space. Dirac's $\delta-$functions $\{ \delta(b_1-x), \delta(b_2-x)\}$ 
form the canonical basis in this space. For each $\varphi \in E$ we have
\[\varphi(x)=\varphi(b_1) \delta(b_1-x) + \varphi(b_2) \delta(b_2-x).\]
By using the representation (\ref{EX1}) we construct the map 
\begin{equation}
\label{MAP}
J^{b/a}:{\cal C} \to \Phi(X, {\bf C})
\end{equation}
The $J^{b/a}$ maps contexts (complexes of, e.g., physical conditions) into complex
amplitudes. The representation ({\ref{Born}}) of probability as the square of the 
absolute value of the complex $(b/a)-$amplitude is nothing other than the 
famous {\bf Born rule.}

{\bf Remark 1.} {\small We underline that the complex linear space representation (\ref{EX1}) of the 
set of contexts ${\cal C}$ is based on a pair $(a,b)$ of incompatible 
(Kolmogorovian) random variables. Here $\varphi_C=\varphi_C^{b/a}.$}

The complex amplitude $\varphi_C(x)$ can be called a {\bf wave function} of the 
complex of physical conditions, context $C,$ cf [14]- [16], of a {\it pure state.} 

We recall that we 
obtained complex probability amplitudes in the conventional Kolmogorov 
framework without appealing to the standard wave or Hilbert space 
arguments. As we shall see, the map $J^{b/a}$ gives a {\bf quantum-like representation} 
of conventional Kolmogorov probability model.

In principle, we can represent each context $C \in {\cal C}$ by a family of complex amplitudes:
\begin{equation}
\label{EX}
\varphi(x)\equiv \varphi_C(x)=\sum_{y\in Y}\sqrt{p_C^a(y)p(x/y)} 
e^{i \xi_C(x/y)}
\end{equation}
such that 
\[\xi_C(x/a_1) - \xi_C (x/a_2)=\theta_C(x) .\]
For such complex amplitudes we also have Born's rule (\ref{Born}).
However, to simplify considerations we shall consider only the representation (\ref{EX1}) and the map 
(\ref{MAP}) induced by this representation.

{\bf 3.2. Hilbert space representation of Born's rule.} We set 
\[e_x^b(\cdot)=\delta(x- \cdot)\]
The representation (\ref{Born}) can be rewritten in the following form:
\begin{equation}
\label{BH}
p_C^b(x)=\vert(\varphi_C, e_x^b)\vert^2 \;,
\end{equation}
where the scalar product in the space $E=\Phi(X, C)$ is defined by the 
standard formula:
\[(\varphi, \psi)=\sum_{x\in X} \varphi(x)\bar \psi(x)\]
The system of functions $\{e_x^b\}_{x\in X}$ is an orthonormal basis in the 
Hilbert space $H=(E, (\cdot, \cdot))$

Let $X \subset R.$ By using the Hilbert space representation of 
Born's rule ({\ref{BH}}) we obtain  the Hilbert space representation of the
expectation of the (Kolmogorovian) random variable $b$:
\begin{equation}
\label{BI1}
E (b/C)= \sum_{x\in X}xp_C^b(x)=\sum_{x\in X}x\vert\varphi_C(x)\vert^2=(\hat b\varphi_C, \varphi_C) \;,
\end{equation}
where $\hat b:\Phi(X,{\bf C})\to \Phi(X, {\bf C})$
is the multiplication operator. This operator can also be determined by its 
eigenvectors: $\hat b e_x^b=x e^b_x, x\in X.$

We set 
$$
u_j^a=\sqrt{p_C^a(a_j)}, u_j^b=\sqrt{p_C^b(b_j)}, p_{ij}=p(b_j/a_i), u_{ij}=\sqrt{p_{ij}},
\theta_j=\theta_C(b_j), \epsilon_j=\epsilon(b_j) \;.
$$
We remark that the coefficients $u_j^a, u_j^b$ depend on a context $C;$ so
$u_j^a=u_j^a(C), u_j^b=u_j^b(C).$
We also consider the {\it matrix of transition 
probabilities} ${\bf P}^{b/a}=(p_{ij}).$
It is always a {\it stochastic matrix.}\footnote{So $p_{i1}+p_{i2}=1, i=1,2.$}
We have, see (\ref{EX}), that
$$
\varphi_C=v_1^b e_1^b + v_2^b e_2^b, \;\mbox{where}\;\;
v_j^b=u_1^a u_{1j}  + u_2^a u_{2j} e^{i\epsilon_j \theta_j}\;.
$$
So
\begin{equation}
\label{BI}
p_C^b(b_j) =\vert v_j^b \vert^2=\vert u_1^a u_{1j}  + u_2^a u_{2j} e^{i\epsilon_j \theta_j}\vert^2.
\end{equation}

This is the {\it interference representation of probabilities} that is used, 
e.g., in quantum formalism.\footnote{ By starting with the general representation (\ref{EX})
we obtain $v_j^b=u_1^a u_{1j} e^{i\xi_{1j}} + u_2^a u_{2j} e^{i\xi_{2j}}$ and the interference
representation $p_C^b(b_j)=\vert v_j^b\vert^2=\vert u_1^a u_{1j} e^{i\xi_{1j}} + u_2^a 
u_{2j} e^{i\xi_{2j}}\vert^2.$} We recall that we obtained (\ref{BI}) starting 
with the interference formula of total probability, (\ref{Two}).

{\bf 3.3. Born's rule and Hilbert space representations.}
We would like to obtain (\ref{BI}) by using the standard quantum procedure, 
namely, transition from the orthonormal basis $\{e_j^b\}$ corresponding the 
$b-$variable to a new basis $\{e_j^a\}$ which corresponds to the $a- 
$variable. Thus we would like to have Born's rule not only in the $b$-representation,
but also in the $a$-representation. As we shall see, we cannot be lucky in the general case.
Starting from two arbitrary incompatible (Kolmogorovian) random variables 
$a$ and $b$ we obtained a complex linear space representation of the 
probabilistic model which is essentially more general than the standard quantum 
representation. In our (more general) linear representation the ``dual 
variable'' $a$ need not be represented by a symmetric operator (matrix) in 
the Hilbert space $H$ generated by the $b$.

For any context $C_0,$ we can represent the $\varphi=\varphi_{C_0}$ in the form:
\begin{equation}
\label{0}
\varphi=u_1^a e_1^a + u_2^a e_2^a,
\end{equation}
where
\begin{equation}
\label{Bas}
e_1^a= (u_{11}, \; \; u_{12})\; \; 
e_2^a= (e^{i\epsilon_1 \theta_1} u_{21}, \; \; e^{i\epsilon_2 \theta_2} u_{22})
\end{equation}
Here $\{e_i^a\}$ is a system of vectors in $E$ corresponding to the $a-$observable.
We suppose that vectors $\{e_i^a\}$ are lineary independent, so $\{e_i^a\}$ 
is a basis in $E.$ We have:
\[e_1^a=v_{11} e_1^b + v_{12} e_2^b, \; \; \; e_2^a=v_{21} e_1^b + v_{22} e_2^b \]
Here $V=(v_{ij})$
is the matrix corresponding to the transformation of complex amplitudes:
$v_{11}=u_{11}, v_{21}=u_{21}$ and $v_{12}=e^{i\epsilon_1 \theta_1} u_{21}, v_{22}=
e^{ii\epsilon_2 \theta_2} u_{22}.$

We would like to find a class of matrixes $V$ such that 
Born's rule (in the Hilbert space form), see (\ref{BH}), holds true also in the $a-$basis:
\[p_C^a(a_j)=\vert(\varphi, e_j^a)\vert^2 \; .\]
By (\ref{0}) we have Born's rule iff $\{e_i^a\}$ was an {\it orthonormal 
basis,} i.e., the  $V$ is a {\it unitary} matrix. Since we study the 
two-dimensional case (i.e., dichotomous random variables), $V\equiv 
V^{b/a}$ is unitary iff the matrix of transition probabilities ${\bf 
P}^{b/a}$ is {\bf double stochastic.}\footnote{So it is stochastic 
and, moreover, $p_{1j} + p_{2j}=1, j=1,2.$}

However, there is some difficulty. In fact, we constructed the $a$-basis 
starting with one fixed context $C_0.$ The basis $e_j^a$ depends on $C_0$
(via the phases $\theta_{C_0}(x)): e_j^a= e_j^a(C_0).$ 
In principle, the validity of Born's rule for the context  $C_0$ 
in the basis $e_j^a(C_0)$ need not imply this rule for any context
$C$ in the same basis $e_j^a(C_0).$ We shall see that 
for double stochastic matrices of transition probabilities (and only such matrices)
we can really construct the $a$-representation starting with some fixed $C_0.$ However,
we should choose signs $\epsilon(x)$ in the representation (\ref{EX1}) in a special way.
We recall that the map $J^{b/a}$ was constructed for fixed signs $\epsilon_1$ and $\epsilon_2;$
so $J^{b/a}= J^{b/a}(\epsilon_1, \epsilon_2).$

We now investigate this problem. We remind that we constructed the matrix $V$ by using
the fixed context $C_0,$ so $V=V(C_0).$
For any $C \in {\cal C},$ we would like to represent the wave function as
\begin{equation}
\label{LUU}
\phi_C= v_1^a(C) e_1^a(C_0) + v_2^a(C) e_2^a(C_0),\;\; \mbox{where}\;\;\; \vert v_j^a(C) \vert^2= p_C^a(a_j).
\end{equation}
It is clear that, for any $C \in {\cal C},$ we can represent the wave function as
$$
\phi_C(b_1) = u_1^a(C) v_{11}(C_0) + e^{i\epsilon_1 [\theta_C(b_1)- \theta_{C_0}(b_1)]} u_2^a(C) v_{12}(C_0)
$$
$$
\phi_C(b_2) = u_1^a(C) v_{21}(C_0) + e^{i\epsilon_1 [\theta_C(b_2)- \theta_{C_0}(b_2)]} u_2^a(C) v_{22}(C_0)
$$
Thus to obtain (\ref{LUU}) we should have:
\begin{equation}
\label{LUU1}
\epsilon_1 [\theta_C(b_1)- \theta_{C_0}(b_1)]= 
\epsilon_2 [\theta_C(b_2)- \theta_{C_0}(b_2]\; \; (\rm{mod} \; 2\pi)
\end{equation}
for any pair of contexts $C_0$ and $C_1.$
Thus 
\begin{equation}
\label{LUU2}
\Delta(C)= \epsilon_1 \theta_C(b_1)- \epsilon_2\theta_C(b_2)= \Delta
\end{equation}
should be a constant $(\rm{mod} \; 2\pi)$ on ${\cal C}.$

{\bf 3.4. The role of the condition of double stochasticity.}

{\bf Lemma 2.} {\it Let $a$ and $b$ be incompatible random variables and let the 
matrix of transition probabilities ${\bf P}^{b/a}$ be double stochastic. 
Then:
\begin{equation}
\label{SW}
\cos \theta_C (b_1)=-\cos \theta_C (b_2)
\end{equation}
for any context $C\in {\cal C}.$}

{\bf Proof.} By Lemma 1 we have:
\[\sum_{x\in X}\cos\theta_C(x)\sqrt{\Pi_{y\in Y}p_c^a(y) p(x/y)}=0\]

But for a double stochastic matrix $(p(x/y))$ we have:
\[\Pi_{y\in Y} p_c^a (a_1) p(b_1/y)=
\Pi_{y\in Y} p_c^a (a_2) p(b_2/y) .\]
Since random variables $a$ and $b$ are incompatible, we have $p(x/y)\not = 
0, x\in X, y\in Y.$ Since $C\in {\cal C}_{\cal A},$ we have $p_C^a(y)\not = 
0, y\in Y.$ We obtain (\ref{SW}).

\medskip

Thus for a double stochastic matrix ${\bP}^{b/a}$ we can choose
\begin{equation}
\label{EO}
\theta_C(b_2)=\pi-\theta_C(b_1)
\end{equation}

{\bf Proposition 1.} {\it {Let the conditions of Lemma 2 hold true. Then the condition 
(\ref{LUU2}) holds true for any Kolmogorov model iff $\epsilon_1= -\epsilon_2$}}

{\bf Proof.} By (\ref{EO}) we obtain:
$$
\Delta(C)= (\epsilon_1 + \epsilon_2)\theta_C(b_1) - \epsilon_2 \pi
$$
\medskip

Let us denote the unit sphere in the Hilbert space $E=\Phi(X, {\bf C})$ 
by the symbol $S.$ The map $J^{b/a}:{\cal C}\to S$ need not be a surjection (injection), 
see examples in section 6. In general the set of pure states corresponding to a Kolmogorovian
model
$$
S_{\cal  C}\equiv S^{b/a}_{\cal  C}
=J^{b/a}({\cal C})
$$
is just a proper subset of the sphere $S.$ The structure of the set of pure states
$S_{\cal C}$ is determined by the Kolmogorov model. 

We remark that for a 
double stochastic matrix ${\bf P}^{b/a}$ (and $\epsilon_1= -\epsilon_2)$
the condition (\ref{LUU2}) 
does not depend on the set ${\cal C}$ (i.e., a Kolmogorov model). Here
always $\Delta= \pi.$  We also remark that, in fact, only double stochastic matrices ${\bf 
P}^{b/a}$ has such a property. By using calculations which have been done 
in the proof of Lemma 1 we obtain the following more general result.

{\bf Lemma 2a.} {\it Let $a$ and $b$ be incompatible random variables. Then 
for any context $C\in {\cal C}$ the following equality holds true:
\begin{equation}
\label{T}
\cos \theta_{C}(b_1)=-k \cos \theta_C (b_2)
\end{equation}
where}
\[k\equiv k^{b/a}=\sqrt{\frac{p_{12}p_{22}}{p_{11}p_{21}}}\]

{\bf Proposition 2.} {\it Let $k>0$ be a real number and let 
angles $\theta_1, \theta_2 \in [0,\pi]$ be connected 
by (\ref{LUU2}). If for all $\theta_2\in [0,\pi]$
$$
\cos\theta_1=-k \cos \theta_2,
$$ 
then $k=1$ and $\Delta=\pi.$ }

{\bf Proof.} By (\ref{LUU2}) we have $\theta_1= \epsilon_1 \Delta+ \epsilon_1\epsilon_2\theta_2.$
Thus $\cos (\epsilon_1 \Delta+ \epsilon_1\epsilon_2\theta_2) = -k \cos \theta_2$ for all
$\theta_2 \in [0, \pi].$ So $\cos (\Delta+ \epsilon_2 \theta_2) = -k \cos \theta_2.$
Let $\theta_2 = \epsilon_2 ( -\Delta + \pi/2).$ So $\cos (- \Delta+ \pi/2)=0.$
Thus $\Delta=0$ or $\Delta= \pi.$ Let $\Delta=0.$ Then $\cos\theta =-k \cos \theta$ for 
any $\theta \in [0, \pi].$ This contradicts to positivity of $k.$ So $\Delta= \pi$ and $k=1.$ 
To get both $\theta_1, \theta_2 \in [0, \pi]$ we should choose $\epsilon_1= -\epsilon_2.$

We also remark that $k^{b/a}=1$ iff ${\bf P}^{b/a}$ is double 
stochastic.

{\bf 3.5. Extension of the Hilbert space representation map.}
The sets $A_i$ are not contexts with respect to $\cal A,$
since ${\bf P}(A_1 A_2)=0$. 
Thus $J^{b/a}(A_i)$ cannot be defined by (\ref{EX1}). 
It is natural to extend the map  $J^{b/a}$ to sets $A_i$ 
by setting 
$$
J^{b/a}(A_i)=e_i^a, i=1,2.
$$
We set 
$$
\overline{{\cal C}}={\cal C} \cup {\cal A}.
$$ 
Thus we have constructed the Hilbert space representation:
\[J^{b/a}:\overline{\cal C}\to S\]
We set $S_{\overline{\cal C}}=J^{b/a}\overline{\cal C}.$

{\bf 3.6. Nonsensitive contexts.} Let $\delta(B_i/{\cal A}, C)=0, i=1,2.$ 
So $\lambda(B_i/{\cal A}, C)=0$ and, hence, $\theta(B_i/{\cal A}, C)= \pi/2.$ Here
(for $x \in X):$
\begin{equation}
\label{Z}
\varphi_C(x)=J^{b/a}(C)(x)=
\sqrt{p_C^a(a_1) p(x/a_1)} + e^{i\epsilon(x)\frac{\pi}{2}} \sqrt{p_C^a(a_2) p(x/a_2)}
\end{equation}
Thus 
\begin{equation}
\label{LUT}
\varphi_C(x)= \sqrt{p_C^a(a_1) p(x/a_1)} + \epsilon(x) i \sqrt{p_C^a(a_2) p(x/a_2)}
\end{equation}

We set 
$$
{\cal C}_0= \{ C \in {\cal C}: \delta(B_j/{\cal A}, C)=0\}.
$$
Contexts $C\in {\cal C}_0$ are said to be $b/a$-nonsensitive contexts. These are  complexes of physical 
(or, e.g., social) conditions $C$ such that a measurement of $a$ under $C$ does not disturb 
the probability distribution of $b.$ We remark that $\Omega$ always belong to ${\cal C}_0.$
However, in general ${\cal C}_0\not= \{ \Omega\},$ see section 6.

{\bf 3.7. Non injectivity of the Hilbert space representation map.}
Let $C_1, C_2 \in {\cal C}$ be contexts such that probability distributions
of random variables $a$ and $b$ under $C_1$ and $C_2,$ respectively, coincide:
$$
p_{C_1}^a(y)= p_{C_2}^a(y), y \in Y, \; \; 
p_{C_1}^b(x)=p_{C_2}^b(x), x \in X.
$$
In such a case $\delta(b=x/ {\cal A}, C_1)= \delta(b=x/ {\cal A}, C_2).$
Thus corresponding phases also coincide: 
$\theta(b=x/ {\cal A}, C_1)= \theta(b=x/ {\cal A}, C_2).$
Hence $\phi_{C_1}(x) = \phi_{C_2}(x), x \in X,$  and
$J^{b/a}(C_1)= J^{b/a}(C_2),$ see section 6 for examples.

{\bf 3.8. Nonquantum Hilbert space representations of Kolmogorovian
models.} Of course, for arbitrary random variables $a$ and $b$ the matrix ${\bf 
P}^{b/a}$ need not be double stochastic. Thus a representation of 
probabilities by vectors in a {\it single Hilbert space} we can obtain for a very 
restricted class of random variables. In particular, such random variables 
are considered in quantum theory (in the formalism of Dirac-von Neumann). In 
general, for each random variable we should introduce its own scalar 
product and corresponding Hilbert space:

\medskip

$H_b=(E, (\cdot, \cdot)_b), H_a=(E, (\cdot, \cdot)_a), \ldots, $ where 
$$
(\varphi, \psi)_b=\sum_{j}v_j^b \bar w_j^b\;\mbox{for}\; \;  
\varphi=\sum_j v_j^b e_j^b, \psi=\sum_j w_j e_j^b,
$$ 
and 
$$
(\varphi, \psi)_a =\sum_j v_j^a \bar w_j^a \; \mbox{for} \; \varphi=\sum_j v_j^a e_j^a, \psi=\sum_j w_j^a e_j^a.
$$ 
The Hilbert spaces 
$H_b, H_a,...$ give the $b-$representation, the $a-$representation, $\ldots.$
Thus $p_C^b(b_j)=\vert(\varphi, e_j^b)_b\vert^2$ and $p_C^a(a_j) 
=\vert(\varphi, e_j^a)_a\vert^2$ and so on. In the $H_a$ we have:
\[E(a/C) = \sum_{y\in Y} y p_C^a(y)=a_1\vert(\varphi_C, e_1^a)_a\vert^2 + 
a_2\vert(\varphi_C, e_2^a)_a\vert^2=(\hat {a}\varphi_C, \varphi_C)_a \;,\]
where the operator $\hat{a} :E\to E$ is determined by its eigenvectors: $\hat{a} 
e_j^a=a_j e_j^a.$

Of course, the representation of random variables by linear operators is 
just a convenient mathematical tool to represent the average of a random 
variable by using only the Hilbert space structure. We recall that we started
with purely ``classical'' Kolmogorovian random variables.

As in the conventional quantum formalism we can also consider the map
\begin{equation}
\label{MAP1}
\tilde{J}^{b/a}:\bar{\cal C} \to \tilde{\Phi}(X, {\bf C}) .
\end{equation}
Here $\tilde{\Phi}(X, {\bf C})$ is the space of equivalent classes of 
functions under the equivalence relation: $\varphi$ equivalent   $\psi$ iff 
$\varphi= t\psi, t \in {\bf C,} \vert t\vert=1,$ and
$\tilde{J}^{b/a}(C)= t \phi_C, t \in {\bf C,} \vert t\vert=1,$ where $C\in \bar{\cal C}.$

\medskip

{\bf{Conclusion.}} {\it In the contextual probabilistic approach we can construct a natural map 
from the set of contexts into the unit sphere of the complex Hilbert space. 
Such a map is determined by a pair $a, b$ of incompatible random variables. 
Unitarity of the matrix $V^{b/a}$ of transition from the basis $\{e^a_i\}$ to 
the basic $\{e_i^b\}$ (these basis correspond to random variables $a$ and 
$b,$ respectively) is equivalent
to the possibility of using Born's rule both in the $a$ and $b$ 
representations. In general (i.e., for an arbitrary
set of contexts) such a construction can be realized only for {\bf double stochastic matrix}
of transition probabilities.}

Everywhere below  we restrict our considerations to the case in which the matrix of transition probabilities
${\bf P}^{b/a}$ is {\bf double stochastic.}

\section{Noncommutativity of operators representing Kolmogorovian random variables}

We consider in this section the case of real valued random variables. Here
spectra  of random variables $b$ and $a$ are subsets of ${\bf R}.$ 

We set $q_1= \sqrt{p_{11}}=\sqrt{p_{22}}$ and $q_2= \sqrt{p_{12}}= \sqrt{p_{21}}.$

Thus the vectors of the $a$-basis, see (\ref{Bas}), have the following form:
$$
e_1^a= (q_1, q_2), \; \; e_2^a= (e^{i \epsilon_1 \theta_1} q_2, e^{i \epsilon_2 \theta_2} q_1)\;.
$$
Since $\theta_1+ \theta_2 = \pi,$ we get $e_2^a= e^{i \epsilon_2 \theta_2} (- q_2,  q_1).$
The factor $e^{i \epsilon_2 \theta_2}$ does not play any role in probabilistic considerations.
Hence we can work in the new basis:
$$
e_1^a= (q_1, q_2), \; \; e_2^a= (-q_2, q_1).
$$
We now find matrices of operators $\hat{a}$ and $\hat{b}$ in the $b$-representation. The latter one
is diagonal. For $\hat{a}$ we have:
$\hat{a}= V \rm{diag}(a_1, a_2) V^\star,$ where $v_{11}=v_{22}=q_1, v_{21}=-v_{12}=q_2.$ Thus
$$
a_{11}= a_1q_1^2 +a_2 q_2^2, \; a_{22}= a_1q_2^2 +a_2 q_1^2,\;
a_{12} = a_{21}= (a_1-a_2) q_1 q_2 .
$$
Hence 
$$
[\hat{b}, \hat{a}] = \hat{m},
$$
where $m_{11}=m_{22}=0$ and $m_{12}=- m_{21}= (a_1- a_2) (b_2-b_1) q_1 q_2.$
Since $a_1\not= a_2, b_1\not= b_2$ and $q_j\not=0,$ we have $\hat{m}\not=0.$

\section{The role of simultaneous double stochasticity of ${\bf P}^{b/a}$ and 
${\bf P}^{a/b}$}

Starting with the $b$-representation -- complex amplitudes $\phi_C(x)$ defined on
the spectrum (range of values) of a random variable $b$ -- we
constructed the $a$-representation. This construction is natural (i.e., reproduce
Born's probability rule) only in the case in which ${\bf P}^{b/a}$ is double stochastic.
We would like to have a symmetric model. So by starting with
the $a$-representation -- complex amplitudes $\phi_C(y)$ defined on
the spectrum (range of values) of a random variable $a$ -- we would like
to construct the natural $b$-representation. Thus both matrices of transition
probabilities ${\bf P}^{b/a}$ and ${\bf P}^{a/b}$ should be double stochastic.

{\bf Theorem 2.} {\it Let the matrix ${\bf P}^{b/a}$ be double stochastic. The contexts
$B_1, B_2$ belong to ${\cal C}$ iff the matrix ${\bf P}^{a/b}$ is double stochastic.}

{\bf Proof.} We have 
$$
\lambda(B_2/{\cal A}, B_1) = -\frac{\mu_1^2 +\mu_2^2}{2\mu_1 \mu_2},
$$
where $\mu_j= \sqrt{p^a_{B_1}(a_j) p(b_2/a_j)}.$ So $\lambda(B_2/{\cal A}, B_1)\geq 1$
and we have the trigonometric behaviour only in the case $\mu_1= \mu_2.$ Thus:
$
p^a_{B_1}(a_1) p(b_2/a_1)=p^a_{B_1}(a_2) p(b_2/a_2).
$
In this case $\lambda(B_2/{\cal A}, B_1)= -1,$ so $\theta(B_2/{\cal A}, B_1)= \pi,$ 
and consequently $\theta(B_1/{\cal A}, B_1)=0.$ We pay attention to the fact that
$p^a_{B_i}(a_j)= p^{a/b}(a_j/b_i)\equiv p(a_j/b_i).$ Thus we have:
\begin{equation}
\label{EQW}
p(a_1/b_1) p(b_2/a_1)= p(a_2/b_1) p(b_2/a_2).
\end{equation}
In the same way by using conditioning with respect to $B_2$ we obtain:
$$
p(a_1/b_2) p(b_1/a_1)= p(a_2/b_2) p(b_1/a_2).
$$
By using double stochasticity of ${\bf P}^{b/a}$ we can rewrite the last
equality as
\begin{equation}
\label{EQW1}
p(a_1/b_2) p(b_2/a_2)= p(a_2/b_2) p(b_2/a_1).
\end{equation}
Thus by (\ref{EQW}) and (\ref{EQW1}) we have:
$$
\frac{p(a_1/b_2)}{p(a_2/b_1)} =\frac{p(a_2/b_2)}{p(a_1/b_1)}.
$$
Hence $p(a_1/b_2) = t p(a_2/b_1)$ and $p(a_2/b_2)= t p(a_1/b_1), t >0.$ 
But $1= p(a_1/b_2) +p(a_2/b_2)= t[ p(a_2/b_1) + p(a_1/b_1)] =t.$

To finish the proof we need the following well known result:

{\bf Lemma 3.} {\it Both matrices of transition probabilities ${\bf P}^{b/a}$ and ${\bf P}^{a/b}$ 
are double stochastic iff the transition probabilities are symmetric, i.e.,
\begin{equation}
\label{SYM}
p(b_i/a_j)=p(a_j/b_i), i, j=1,2 .
\end{equation}
This is equivalent that random variables $a$ and $b$ have the uniform probability distribution:
$p^a(a_i)=p^b(b_i)=1/2, i=1,2.$}

This Lemma has important physical consequences. A natural (Bornian) Hilbert space representation
of contexts can be constructed only on the basis of a pair of (incompatible) uniformly distributed
random variables.

{\bf Lemma 4.} 
{\it{Let both matrices ${\bf P}^{b/a}$ and ${\bf P}^{a/b}$ be double stochastic. Then}}
\begin{equation}
\label{L}
\la (B_i/{\cal A}, B_i) =1 .
\end{equation}

\noindent
{\bf Proof.} Here $\delta(B_i/{\cal A}, B_i)=1-p(b_i/a_1) p(a_1/b_i)-p(b_i/a_2) p(a_2/b_i)
=1-p(a_1/b_i)^2- p(a_2/b_i)^2 = 2p(a_1/b_i) p(a_2/b_i).$ Thus 
$
\lambda(B_i/{\cal A}, B_i)=1.
$
\bigskip

By (\ref{L}) we have \[\la (B_i/{\cal A}, B_j)=-1, i \not = j,\]
Thus
\[\theta (B_i/{\cal A}, B_i)=0\; \mbox{and} \;\theta (B_i/{\cal A}, B_j)=\pi, i \not = j.\]

\noindent

{\bf Proposition 2.} 
{\it{Let both matrices of transition probabilities ${\bf P}^{b/a}$ and ${\bf P}^{a/b}$ be double stochastic. Then}}
$$
J^{b/a}(B_j)(x)=\delta(b_j-x), x  \in X, \;\; \mbox{and} \; \; \; J^{a/b}(A_j)(y)=\delta(a_j-y), y \in Y.
$$
\noindent
{\bf Proof.} 
Because $\theta(B_1/{\cal A,} B_1)=0$ we have: 
$$
J^{b/a}(B_1)(b_1)=
\sqrt{p(a_1/b_1) p(b_1/a_1)} + e^{i0} \sqrt{p(a_2/b_1) p(b_1/a_2)}
$$
$$
= 
p(a_1/b_1)+ p(a_2/b_1)=1.
$$ 
Because $\theta (B_2/{\cal A,} B_1)=\pi$ we have 
$$
J^{b/a}(B_1)(b_2)=\sqrt{p(a_1/b_1) p(b_2/a_1)} 
+ e^{i\pi} \sqrt{p(a_2/b_1) p(b_2/a_2)}
$$
$$
=
\sqrt{p(a_1/b_1)}(\sqrt{p(b_2/a_1}-\sqrt{p(a_2/b_1)})=0.
$$
Thus in this case:
\[J^{b/a}(B_i)=e^b_i ,  i=1,2.\]

\section{Example of the Hilbert space representation of the contextual Kolmogorovian model}

We consider an example of a Kolmogorov probability space and a pair of dichotomous random
variables $a, b$ which are incompatible. In this example 
the set of contexts with nontrivial disturbance term $\delta, \delta \not = 0,$ is nonempty,
so ${\cal C}_0 \not= {\cal C}.$

{\bf 6.1. Kolmogorov probability space and incompatible random variables.}
We find the image $S_{\cal C}$ of the set of contexts $\cal C$ in the Hilbert sphere 
$S \subset E=\Phi(X, C).$ In this example $S_{\cal C}$ is a proper subset
of the sphere $S.$ The Hilbert space representation map $J^{b/a}$ is not injective.
Random variables $a$ and $b$ are represented by symmetric 
operators in the Hilbert space $E.$ They do not commute. 

Let $\Omega=\{\omega_1, \omega_2, \omega_3, \omega_4\}$ and ${\bf P}(\om_j)=p_j>0, \sum_{j=1}^4 p_j=1.$ Let 
\[A_1=\{\omega_1, \omega_2\}, A_2=\{\omega_3, \omega_4\}\]
\[B_1=\{\omega_1, \omega_4\}, B_2=\{\omega_2, \omega_3\}\]

Let $p_1=p_3=q < \frac{1}{2}$ and $p_2=p_4=(1-2q)/2.$ We denote this Kolmogorov
probability space by the symbol ${\cal K}(q).$

Here ${\bf P}(A_1)={\bf P}(A_2)={\bf P}(B_1)={\bf P}(B_2)=\frac{1}{2}.$ So the random 
variables $a$ and $b$ are uniformly distributed. Thus both matrices of transition probabilities
${\bf P}^{b/a}$ and ${\bf P}^{a/b}$ are double stochastic. Here 
\[{\bf P}^{b/a}={\bf P}^{a/b}= \left( \begin{array}{lr}
2q  & 1-2q\\
1-2q & 2q
\end{array}
\right) \]
We have the symmetry condition ${\bf P}(B_i/A_j)={\bf P}(A_j/B_i).$

{\bf 6.2. Hilbert space representation of contexts.} We choose $\epsilon_1=-1$ and $\epsilon_2=+1$ 
to fix the map $J^{b/a}.$
We start with two-points contexts. 

(a) Let $C=C_{24}=\{\omega_2, \omega_4\}.$ Here ${\bf P}(C)=1-2q, {\bf P}(B_j/C)={\bf P}(A_j/C)=\frac{1}{2}.$ 
Thus $\delta=0.$ By using the representation (\ref{LUT}), we obtain:
\begin{equation}
\label{T_1}
\varphi_{C_{24}}(x)= \left\{ \begin{array}{ll}
{{\sqrt{q}}-i \sqrt{\frac{1-2q}{2}}, x=b_1}\\
{\sqrt{\frac{1-2q}{2}}+i\sqrt{q}, x=b_2}
\end{array}
\right .
\end{equation}

(b). Let $C=C_{13}=\{\omega_1, \omega_3\}.$ Here everything is as in (a). 
So we have $\varphi_{C_{13}}=\varphi_{C_{24}}$
Thus $J^{b/a}$ is not injective: $J^{b/a}(C_{24}) =J^{b/a}(C_{13}).$

(c) Let $C=C_{14}=\{\omega_1, \omega_4\}=B_1.$ By general theory we have 
$\varphi_{C_{14}}(x)=\delta(b_1-x)=e_1^b.$ 
In the same way: $\varphi_{C_{23}}=\delta(b_2-x)=e_2^b.$

To find the Hilbert space representation of sets $C=C_{12}=\{\omega_1, \omega_2\}=A_1$ and
$C=C_{34}=\{\omega_3, \omega_4\}=A_2$ we have to construct the basis $\{e_j^a\}.$ 
We can choose:
\[e_1^a= \left( \begin{array}{cc}
{\sqrt{2q}}\\
{\sqrt{1-2q}}
\end{array}
\right)
\; \; e_2^a= \left( \begin{array}{cc}
{-\sqrt{1-2q}}\\
{\sqrt{2q}}
\end{array}
\right)\]
(d) Let $C=C_{123}=\{\omega_1, \omega_2, \omega_3\}.$ Here ${\bf P}(C)=(2q+1)/2,
{\bf P}(A_1/C)={\bf P}(B_2/C)=1/(2q+1), {\bf P}(A_2/C)={\bf P}(B_1/C)=2q/(2q+1).$ 
Thus $\delta(B_1/{\cal A}, C)=\frac{2q(2q-1)}{2q+1}$ and, 
hence, $\la(B_1/{\cal A}, C)=-\frac{\sqrt{1-2q}}{2}.$ This context is trigonometric, i.e., 
the measurement of the random variable $a$ under the complex of physical conditions $C$ induces 
 nontrivial, but relatively small statistical disturbance of the "$b$-property"; so 
$C_{123}\in {\cal C}.$ We remark that $\la(B_2/{\cal A}, C)=\frac{\sqrt{1-2q}}{2}$ 
(since ${\bf P}^{b/a}$ is double stochastic).\footnote{ 
We pay attention on the dependence of $\theta= \arccos \frac{\sqrt{1-2q}}{2}$ 
on the parameter $q:\theta(q)$ increases from $\pi/3$ to $\pi/2,$ 
when $q$ increases from 0 to 1/2.} We have:

\[\varphi_{C_{123}}(x)= \left\{ \begin{array}{ll}
{\sqrt{\frac{2q}{2q+1}}-e^{i \arccos \frac{\sqrt{1-2q}}{2}}\sqrt{\frac{2q(1-2q)}{2q+1}}, \;x=b_1}\\
{\sqrt{\frac{1-2q}{2q+1}}+e^{i \arccos \frac{\sqrt{1-2q}}{2}}{\frac{2q}{\sqrt{2q+1}}}, \;\;\;\;\;\;x=b_2}
\end{array}
\right. \]
{\bf Remark.} In principle, we could choose, e.g.,
\[e_2^a= \left( \begin{array}{cc}
{-e^{i\theta}\sqrt{1-2q}}\\
{e^{i\theta}\sqrt{2q}}
\end{array}
\right)
, \; \; \theta=\arccos \frac{\sqrt{1-2q}}{2}.\]
Thus
 \[\varphi_{C_{123}}=\frac{1}{\sqrt{2q+1}}e_1^a + e^{i\arccos \frac{\sqrt{1-2q}}{2}}\sqrt{\frac{2q}{2q+1}}e_2^a\;.\]

\noindent
(e) Let $C=C_{124}=\{\omega_1, \omega_2, \omega_4\}.$ 
Here ${\bf P}(C)=1-q, {\bf P}(A_1/C)={\bf P}(B_1/C)=1/2(1-q), {\bf P}(A_2/C)={\bf P}(B_2/C)=(1-2q)/2(1-q).$ 
Thus $\delta(B_1/{\cal A}, C)=q(1-2q)/(1-q)$ and, hence, $\la(B_1/{\cal A}, C)=\sqrt{\frac{q}{2}}<1,$  
and the context $C_{124} \in {\cal C}.$ Thus:

\[\varphi_{C_{124}}(x)= \left\{ \begin{array}{ll}
{\sqrt{\frac{q}{1-q}}+      e^{-i \arccos \sqrt{\frac{q}{2}}}\frac{1-2q}{\sqrt{2(1-q)}},\;\;\;\;\; x=b_1}\\
{\sqrt{\frac{1-2q}{2(1-q)}}-e^{-i \arccos \sqrt{\frac{q}{2}}}\sqrt{\frac{q(1-2q)}{1-q}},\; x=b_2}
\end{array}
\right . \]
\[\varphi_{C_{124}}(x)=
\frac{1}{\sqrt{2(1-q)}} \;e_1^a -e^{-i \arccos \sqrt{\frac{q}{2}}}\sqrt{\frac{1-2q}{2(1-q)}}\; e_2^a.\]

(f) Let $C=C_{234}=\{\omega_2, \omega_3, \omega_4\}.$ 
Here ${\bf P}(C)=1-q, {\bf P}(A_1/C)={\bf P}(B_1/C)=(1-2q)/2(1-q), {\bf P}(A_2/C)={\bf P}(B_2/C)=1/2(1-q).$ 
Thus $\delta(B_1/{\cal A}, C)=q(2q-1)/(1-q)$ 
and, hence, $\la(B_1/{\cal A}, C)=-\sqrt{\frac{q}{2}},\la(B_2/{\cal A}, C)=\sqrt{\frac{q}{2}}.$ 
Here:
\[\varphi_{C_{234}}(x)= \left\{ \begin{array}{ll}
{\sqrt{\frac{q(1-2q)}{1-q}}-e^{i \arccos \sqrt{\frac{q}{2}}}\sqrt{\frac{1-2q}{2(1-q)}}, \;x=b_1}\\
{\frac{1-2q}{\sqrt{2(1-q)}}+e^{i \arccos \sqrt{\frac{q}{2}}}\sqrt{\frac{q}{1-q}}, \;\;\;\;\;\;x=b_2}
\end{array}
\right. \]
\[\varphi_{C_{234}}(x)=\sqrt{\frac{1-2q}{2(1-q)}}\; e_1^a +e^{i \arccos \sqrt{\frac{q}{2}}}\frac{1}{\sqrt{2(1-q)}} \;e_2^a\;.\]

(g) Let $C=C_{134}=\{\omega_1, \omega_3, \omega_4\}.$ 
Here ${\bf P}(C)=(2q+1)/2, {\bf P}(A_1/C)={\bf P}(B_2/C)=2q/(2q+1), {\bf P}(A_2/C)={\bf P}(B_1/C)=1/(2q+1).$ 
Thus $\delta(B_1/{\cal A}, C)=2q(1-2q)/(2q+1)$ 
and, hence, $\la(B_1/{\cal A}, C)=\frac{\sqrt{1-2q}}{2}.$ 
Thus:
\[\varphi_{C_{134}}(x)= \left\{ \begin{array}{ll}
      {\frac{2q}{\sqrt{2q+1}}+e^{-i \arccos \frac{\sqrt{1-2q}}{2}}\sqrt{\frac{1-2q}{2q+1}}, \;\;\;\;\;x=b_1}\\
{\sqrt{\frac{2q(1-2q)}{2q+1}}-e^{-i \arccos \frac{\sqrt{1-2q}}{2}}\sqrt{\frac{2q}{2q+1}}, x=b_2}
\end{array}
\right. \]

\[\varphi_{C_{134}}=\sqrt{\frac{2q}{2q+1}} \;e_1^a -e^{-i \arccos \frac{\sqrt{1-q}}{2}}\frac{1}{\sqrt{2q+1}} \;e_2^a.\]

\noindent
(h) Let $C=\Omega.$ Here we know from the beginning that $\delta(B_j/{\cal A}, C)=0.$ 
Here ${\bf P}(A_i/C)={\bf P}(A_i)=1/2$ and ${\bf P}(B_i/C)={\bf P}(B_i)=1/2.$ 
Thus $J^{b/a}(\Omega) = J^{b/a}(C_{24}) =J^{b/a}(C_{13})=$(\ref{T_1}).

In this example the set of nonsensitive contexts contains three contexts: ${\cal C}_0=
\{ \Omega, C_{24}, C_{13}\}.$ We have 
$$
S_{\bar{\cal C}}= \{\varphi_\Omega, \varphi_{C_{14}}=e_1^b, \varphi_{C_{23}}=e_2^b, 
\varphi_{C_{12}}=e_1^a, \varphi_{C_{23}}=e_2^a, 
\varphi_{C_{124}}, \varphi_{C_{234}}, \varphi_{C_{123}}, \varphi_{C_{134}}\}
$$
So the set of pure states $S_{\bar{\cal C}}$ is a finite, five-points, subset of the unit sphere
in the two dimensional Hilbert space.

We remark that there is a parameter $q \in (0, 1/2)$ determining 
a Kolmogorov probability model ${\cal K}(q).$ For each value of $q$ we have  a finite set of pure states.
However, a family ${\cal K}(q), q\in (0, 1/2),$ of Kolmogorov probability spaces generates a ``continuous'' set 
$\cup_q S_{\bar{\cal C}}(q)$ of pure states.

\section{Contextual correspondence between Kolmogorovian random variables and 
quantum observables}

We begin with the following standard definition:
 
 {\bf Definition 3.} {\it For a self-adjoint operator $\hat d$ the quantum mean value in the state $\varphi$ is defined by}
 \[\langle \hat d \rangle_\varphi= (\hat d \varphi, \varphi).\]
 
 {\bf Theorem 3.} {\it{For any map $f:{\bf R} \to {\bf R}$ we have:}}
 \[\langle f(\hat a)\rangle_{\varphi_C}=E(f(a)/C), \;\;\; \langle f(\hat b)\rangle_{\varphi_C}=E(f(b)/C)\]
 for any context $C \in \bar {\cal C}.$
 
 {\bf Proof.} By using Borness of the $b$-representation we obtain:
 \[E(f(b/C)=\sum_{x\in X} f(x) p_c^b(x)=\sum_{x \in X} f(x)|(\varphi_C, e_x^b)|^2=\langle f(\hat b)\rangle_{\varphi_C}\]
The same result we have for the $f(\hat a)$ since (as ${\bf P}^{b/a}$ is double stochastic)
we have Born's probability rule both for $b$ and $a.$

{\bf Theorem 4.} {\it Let $f, g: {\bf R} \to {\bf R}$ be two arbitrary functions. 
Then 
\[E(f(a) + g(b)/C)=\langle f(\hat a) + g(\hat b)\rangle_{\varphi_C}\] 
for any context $C \in {\bar {\cal C}}.$}

{\bf Proof.} By using linearity of the Kolmogorov mathematical expectation, Theorem 3,
and linearity of the Hilbert space scalar product we obtain:
\[E(f(a(\omega)) + g(b(\omega))/C)=E(f(a(\omega)/C) + E(g(b(\omega))/C)\]
\[= \langle f(\hat a)\rangle_{\varphi_C} + \langle g(\hat b)\rangle_{\varphi_C}= 
\langle f(\hat a) + g(\hat b)\rangle_{\varphi_C}\]

Denote the linear space of all random variables of the form 
$d(\omega)=f(a(\omega)) + g(b(\omega))$ by the symbol ${\cal O}_+(a,b)$ 
and the linear space of operators of the form $\hat d=f(\hat a) + g(\hat b)$ by 
${\cal O}_+(\hat a, \hat b).$

{\bf Theorem 5.} {\it  The map $T=T^{a/b}:{\cal O}_+(a, b) \to {\cal O}_+(\hat a, \hat b),
d=f(a) + g(b) \to \hat d=f(\hat a) + g(\hat b),$ 
preserves the conditional expectation:}
\begin{equation}
\label{R}
\langle T(d)\rangle_{\varphi_C}=(T(d) J(C), J(C))=E(d/C).
\end{equation}

The transformation $T$ preserves the conditional expectation for random variables $d \in {\cal O}_+(a, b).$
  But in general we cannot expect anything more, since in general $T$ does not preserve
  probability distributions. The important problem is to
  extend the map $T$ for a larger class (linear space?) of Kolmogorovian random variables 
  with preserving (\ref{R}). It is natural  to define (as we always do in the conventional quantum formalism):
  $$
  T(f)(\hat a, \hat b)=f(\hat a, \hat b)
  $$
  where $f(\hat a, \hat b)$ is the pseudo differential operator with the Weyl symbol $f(a, b)$. 
  We shall see that already for $f(a, b)=ab$ (so $f(\hat a, \hat b)=(\hat a \hat b + \hat b \hat a)/2$) 
  the equality (\ref{R}) is violated.

We can consider the $b$ and the $a$ as discrete analogues of the position and momentum observables. 
The operators $\hat b$ and $\hat a$ give the Hilbert space (quantum) representation of these observables.

We also introduce an analogue of the energy observable:
$$
{\cal H}(\omega)=\frac{h}{2}[a^2(\omega) + V(b(\omega))],
$$
where $h>0$ is a constant and $V:{\bf R} \to {\bf R}$ is a map. 
The Hilbert space representation of this observable is given by the operator of energy (Hamiltonian) 
$$
\hat {\cal H} =\frac{h}{2}(\hat a^2 + V(\hat b)).
$$
By Theorem 5 for contexts $C \in \bar{\cal C}$ the averages of the observables ${\cal H}(\omega)$ 
(Kolmogorovian) and $\hat {\cal H}$ (quantum) coincide:
$$
E({\cal H} (\omega)/C) = \langle {\cal H} \rangle_{\varphi_C} .
$$
However, as we shall see, probability distributions do not coincide:

{\bf Proposition 3.} {\it There exists  context $C$ such that the probability distribution 
of the random  variable $d(\omega)=a(\omega) + b(\omega)$ with respect to $C$ does not 
coincide with the probability distribution of the corresponding quantum observable 
$\hat d=\hat a + \hat b$ with respect to the state $\varphi_C$.}

{\bf Proof.} It suffices to present an example of such a context $C.$ 
Take the context $C=C_{234}$ from section 6. We consider the case: 
$a(\omega)=\pm \gamma, b(\omega)=\pm \gamma, \gamma > 0;$ so 
$d(\omega)= - 2\gamma, 0, 2\gamma.$ 
Corresponding Kolmogorovian probabilities can easily be found:
\[p_C^d(-2\gamma)= q/(1-q),\; \;  p_C^d(0)=(1-2q)/(1-q),\; \;  p_C^d(2\gamma)=0.\]
We now find the probability distribution of $\hat d$.
To do this, we find eigenvalues and eigenvectors of the self-adjoint operator
$\hat d.$ 
We find the matrix of the operator $\hat d$ in the basis
$\{ e_j^b \}:$
$ d_{11} = - d_{22}= 4q \gamma $ and $d_{12} = d_{21}=2 \gamma \sqrt{2q(1-2q)}.$
We have $k_{1,2}=\pm 2 \sqrt{2q}\gamma.$
Of course, the range of 
values of the quantum observable $\hat d$ differs from the range of values of the random variable $d.$ 
However, this difference of ranges of values is not so large problem in this case. The random variable $d$ 
takes only two values, $-2\gamma, 0$ with the probability one. Moreover, we can represent values of 
the quantum observable $\hat d$ as just an affine transform of values of the random variable $d:$
\[d_{\rm{quantum}}=2\sqrt{2q} \; d-\gamma.\]
In principle we can interpret such a transformation as representing some special measurement procedure. 
Thus in this example the problem with spectrum is not crucial. The crucial problem is that $d$ and
$\hat d$ have different probability distributions.

Corresponding eigenvectors are
\[e_1^d=\frac{1}{\sqrt{2(1-\sqrt{2q})}} (-\sqrt{1-2q}, \sqrt{2q}-1)\]
\[e_2^d=\frac{1}{\sqrt{2(1+\sqrt{2q})}} (-\sqrt{1-2q}, \sqrt{2q}+1)\]

Finally, we find (by using the expression for $\varphi_{C_{234}}$ which was found in section 6):
\[p_c^{\hat d}(k_1)=|(\varphi_C, e_1^d)|^2=\frac{(1-\sqrt{2q})(2+\sqrt{2q})}{4(1-q)}\]
\[p_c^{\hat d}(k_2)=|(\varphi_C, e_2^d)|^2=\frac{(1+\sqrt{2q})(2-\sqrt{2q})}{4(1-q)}\]
Thus $d$ and $\hat d$ have essentially different probability distributions.

\section{Dispersion-free states}
{\bf 1. Von Neumann.}
As originally stated by von Neumann, [5] the problem of hidden variables is to find whether
{\it{dispersion free states exist}} in QM. He answered the question in the negative. 
The problem of the existence of dispersion free states as well as von Neumann's solution were 
the subject of great debates. We do not want to go into detail see, e.g., [8], [9]. In our contextual
approach an analogue of this problem can be formulated as 

{\bf Do dispersion free contexts exist?}

The answer is the positive. In the example of section 6. we can take any atom of 
the Kolmogorov probability space ${\cal K}_q,$ e.g., $C=\{\omega_1\}.$ Since, for any random variable
$\xi$ on the Kolmogorov space ${\cal K}_q$, it has a constant value on such a $C$ 
the dispersion of $\xi$ under the context $C$ is equal to zero:

\[D(\xi/C)=E[(\xi-E(\xi/C))^2/C]=0.\]

However, dispersion free contexts do not belong to the system $\bar{\cal C}$ of contexts which
can be mapped by $J^{a/b}$ into the Hilbert space $H.$ On the one hand, our contextual approach gives 
the possibility to have the realist viewpoint to QM.  On the other hand, it does not 
contradict to the
von Neumann as well as other ``no-go" theorems. The mathematical representation of contexts 
(complexes of physical conditions) given by the quantum formalism it too rough to represent dispersion 
free contexts.
	
{\bf 2. Kochen and Specker.}
In the model of Kochen and Specker on the  set ${\cal L}_s$ of self-adjoint operators there was considered 
only the structure of a partial algebra:

{\it{Products and sums are defined only for pairs of commutative operators.}}

A necessary condition for the existence of a hidden variable interpretation is then the existence of
an embedding of the partial algebra ${\cal L}_s$ into a commutative algebra. Kochen and Specker proved 
that such an embedding is impossible for Hilbert spaces of dimension $\geq 3.$ Despite the fact that 
we restricted our contextual considerations to representations of Kolmogorovian spaces based on pairs 
of (incompatible) dichotomous variables, we can compare our approach with Kochen-Specker approach. 
In our contextual approach the commutative algebra should be chosen as the algebra
$RV(\Omega, {\cal F}, {\bf P})$ of random variables. But the whole formulation of the problem of
embedding of ${\cal L}_s$ into $RV(\Omega, {\cal F}, {\bf P})$ is meaningless in our framework.
The use of only the structure of prealgebra does not change anything. The operator 
$\hat d=\hat a + \hat b \in {\cal L}_s$ and it has the natural preimage $d=a+b \in 
RV(\Omega, {\cal F}, {\bf P})$. The problem is that $\hat d$ and $d$ can have different
probability distributions.

Thus from our point of view the dimension of a Hilbert space is not important. Even, as we have seen, in 
the two-dimensional case only a very restricted class of variables $d \in RV(\Omega, {\cal F}, {\bf P})$ can
be mapped into ${\cal L}_s$ with preserving of probability distributions.

\section{Classical and quantum spaces as rough images of fundamental prespace}

Our contextual probabilistic model induces the following picture of physical reality. 

{\bf 1. Prespace and classical space.}
There exists a prespace $\Omega$ which points corresponds to primary (irreducible) states 
of physical systems, {\bf prestates or fundamental physical parameters}. Functions 
$d:\Omega \to {\bf R}^m$ are said to be {\bf preobservables.} The set of all preobservables 
is denoted by the symbol ${\cal O}_p \equiv {\cal O}_p (\Omega).$ We are not able (at least at the moment)
to measure an arbitrary preobservable  $d \in {\cal O}_p.$

Nevertheless, some preobservables can be measured. Suppose that there exists a preobservable
$b$ such that all measurements can be reduced to some measurements of $b,$ cf. L. De Broglie [10]
on the possibility to reduce any measurement to a position measurement. Let $X \subset {\bf R}^m$ be the 
range of values \footnote{ See section 8.3 on some motivations to consider $X$ as a subset of ${\bf R}^m.$} of $b.$ The $X$ is said to be a classical space \footnote{ Of course, in such a model the classical space $X$ depends on the preobservable $X\equiv X(b).$ Thus $X$ is the $b$-image of the prespace $\Omega.$}. Set $B_x=\{\omega \in \Omega: b(\omega)=x\}=b^{-1}(x), x \in X.$

In principle a set $B_x$ could contain millions of points. Dynamics in $X$ is classical dynamics. In our model, classical dynamics is a rough image of dynamics in the prespace $\Omega$ \footnote{ Consider in the example of section 6 the trajectory $\omega_1 \to \omega_2 \to \omega_3 \to \omega_4 \to \omega_1$ in the $\Omega$. In the classical space $X$ this trajectory is represented by $b_1 \to b_1 \to b_2 \to b_2 \to b_1$.}.

{\bf 2. Classical phase space.}
Let $a$ be a preobservable which is incompatible with our fundamental preobservable $b$ (space observable). We denote by $Y \subset {\bf R}^m$ the range of values of the $a.$ The $Y$ is said to be conjugate space to the classical space $X.$ We call the $b$ {\it{position}} and the $a$ {\it{momentum.}} We set $A_y=\{\omega \in \Omega: a(\omega)=y\}=a^{-1}(y), y \in Y.$

Since $A_y$ is not a subset of $B_x$ for any $x \in X,$  the point $y$ cannot 
be used to get finer description of any point $x \in X.$ Thus by using 
values of the momentum we cannot obtain a finer space structure. The variables $b$ and $a$ are really 
incompatible. By fixing the value of, e.g., $a=y_0$ we cannot fix the value of $b=x_0.$ 
It is important for future considerations to notice that sets $A_y B_x$
are not contexts (in the contextual(b,a)-picture). In general
$$
A_y  B_x \not\in \bar{\cal C}.
$$
{\bf Remark.} 
(Nonlocal dependence of incompatible variables at the prespace level). Since, for a fixed $y_0 \in Y,$
we have $A_{y_0} \cap B_x \ne \emptyset$ for any $x \in X,$ a value $y_0$ of the momentum can
be determined only by all values $x \in X$ of the position. Thus on the level of 
the prespace incompatible variables are {\bf{nonlocally dependent}}. 
However, this prespace nonlocality could not be found in classical mechanics, since in the latter 
the finer prespace structure is destroyed by the rough $(x, y)$ encoding.

The space $\Pi=X\times Y \subset {\bf R}^{2m}$ is a classical phase space. Dynamics in the phase space gives
a rough image in the terms of the two incompatible variables of dynamics in the prespace.
The phase space $\Pi$ is a classical contextual $(b, a)$-picture of the prespace $\Omega.$ This picture is 
richer than the pure $b$-space picture. The $\Pi$ contains images of the 
two families of contexts ${\cal A}=\{A_y\}$ and ${\cal B}=\{B_x\}.$

{\bf 3. On homogeneous structure of the classical space.}
In our probabilistic investigations we have seen that the most 
natural choice of incompatible variables corresponds to random variables $a(\omega)$ and 
$b(\omega)$ which are uniformly distributed. On the other hand, the creation of a uniform partition 
of the prespace $\Omega$ is the most natural way to create a rough image $X$ of the prespace -- a classical space.

If a group of cognitive systems have used such a uniform partition of the prespace 
then the corresponding classical space should be homogeneous. This is a reason to assume (as we have done) 
that the classical space $X \subset {\bf R}^m.$ But, of course, we could not deduce the real number structure 
of the classical space only 
on the basis of the fact that fundamental variables should be uniformly distributed.

{\bf 4. Classical statistical mechanics.}
As the next step we can consider statistical mechanics on the classical space $X.$ In such a statistical 
theory from the very beginning we lost the finer statistical structure of the model based on probability 
distributions on the prespace. Functions $u: \Pi \to {\bf R}^q$ are called classical observables. 
The set of classical observables is denoted by the symbol ${\cal O}_c (\Pi).$ 
We shall also use symbols ${\cal O}_c (X)$ and  ${\cal O}_c (Y)$ to denote spaces of classical observables 
depending only on the $b$-position and the $a$-momentum, respectively.

{\bf 5. Quantum mechanics and the Hilbert space representation of prespace contexts.}
Neither classical nor quantum mechanics can describe the individual dynamics in the 
prespace. Of course, such a viewpoint to quantum mechanics contradicts to the so called orthodox 
Copenhagen interpretation by which the wave function describes an individual quantum system.
It seems that our contextual approach to quantum theory is closer to the so called statistical 
(or ensemble) interpretation of quantum mechanics. By the latter a wave function describes
not an individual quantum system but  statistical properties of an 
ensemble of quantum systems, see, e.g., [7]. 

By our {\it contextual interpretation} the wave function has a realist prespace interpretation.
A complex amplitude is nothing than an  image (induced by the contextual formula of total
probability) of a set of fundamental parameters - context. Thus the Hilbert state space $H$
is not less real than the classical real space ${\bf R}^3.$

Observables which probability distributions can be found by using the representation by self-adjoint 
operators in the Hilbert space are called quantum observables. The set of quantum observables is denoted
by the symbol ${\cal O}_q(H).$
Neither classical statistical nor quantum mechanics can provide knowledge about the probability distribution of an arbitrary preobservable. Nevertheless, the quantum theory gives some information about some preobservables, namely 
fundamental preobservable $b$ and $a$ and pre-observables $d$ belonging to the class ${\cal O}_+ (a, b).$ 
Another way to look to the same problem is to say that the quantum 
theory (with our contextual probabilistic interpretation) gives the possibility to represent 
some prespace structures, namely some contexts $C \in {\cal C}$ by vectors of a Hilbert state space.

Neither classical nor quantum mechanics are fundamental theories. They could not give 
information about the point wise structure of the prespace $\Omega.$ But the quantum formalism 
represents some complexes of physical conditions -- domains in the prespace -- which are not represented in
the classical space or phase space. Of course, the quantum formalism also represents classical position
states $x \in X$ by wave functions $\varphi_{B_x}$ (Hilbert states $e_x^b$). Classical states $x \in X$ are 
images of prespace contexts $B_x.$ But the quantum formalism represents also some sets $C \subset \Omega$
which have no classical images (namely, images in $X$ or $\Pi$).

{\bf Example.}
In the example of section 6 we take the set $C=C_{123}=\{\omega_1, \omega_2, \omega_3\}.$ 
Neither $C \subset B_1$ nor $C \subset B_2.$ This prespace domain $C$ can be described neither by 
the position $x=b_1$ nor $x=b_2$. The quantum state $\varphi_C \in S \subset H$ representing this 
domain of the prespace describes the superposition of the two classical states $x=b_1$ and $x=b_2.$
Hence a physical system prepared under the complex physical conditions $C=C_{123}$ is (from the classical 
viewpoint) in the superposition of two different positions. 

{\bf 6. Heisenberg uncertainty principle.}
We now take the context $C=A_y$ for some $y \in Y.$
Here the momentum $a$ has the definite value. But  $A_y \cap B_x \ne \emptyset$ for any $x \in X.$
Hence the state $\varphi_C=e_y^a \in {\cal S} \subset H$ also corresponds to the superposition of
two positions  $x=b_1$ and $x=b_2.$
This is nothing else than (the discrete analogue) the {\bf Heisenberg uncertainty 
principle}. In the same way in any state with the definite position, $\varphi_C=e_x^b, x\in X,$ the 
momentum can not have the definite value.

Thus the Hilbert sphere $S$ contains images of the classical spaces $X$ and $Y$ 
(but not the phase space  $\Pi$, see further considerations), $X \subset S$ and $Y \subset S.$ But the Hilbert
space contains also images of nonclassical domains
$C \in \bar{\cal C}.$ We remark that (depending on the model) only a part of the Hilbert sphere
corresponds to some domains of the prespace. All other quantum states, $\varphi \not\in S_{\bar {\cal C}},$
are just ideal mathematical objects which do no correspond to any context in the prespace.

As was already remarked, the phase space  $\Pi$ 
is not imbedded into the Hilbert sphere $S$, since contexts $C_{xy}=B_x  A_y$
corresponding to points of the $\Pi$ do not belong to the system $\bar {\cal C}$ which is mapped
into $S.$

{\bf 7. Preobservables and quantum observables.}
For what class of preobservables can we find probability distributions with respect to contexts 
$C \in \bar{\cal C}$ by using the quantum formalism? As we have seen, we are not able to find the 
probability distribution for an arbitrary $d\in {\cal O}_p(\Omega).$
In general the operators $\hat d=d(\hat a, \hat b)$ corresponding to functions 
$d(x, y)$ (e.g., $d(x, y)=xy$ or $d(x, y)=x+y)$
are not directly related to prequantum observables $d(\omega)=d(b(\omega), a(\omega)).$

Only quantum observables $\hat d=f(\hat b)$ and $\hat d=g (\hat a)$ have the same probability
distributions as the corresponding preobservables $d(\omega)=f(b(\omega))$ and $d(\omega)=g(a(\omega)).$
By Theorem? the average is preserved by the canonical map $T^{b/a}: {\cal O}_+(a, b) \to 
{\cal O}_+(\hat a, \hat b).$

However, even such quantum observables give just a rough image of corresponding preobservables. 
By using quantum probabilistic formalism we can find probability distributions only for quantum 
states $\varphi_C \in S_{\bar{\cal C}} \subset H$. 
Those quantum states represent only some special contexts. Hence by using the quantum formalism 
we could not find the probability distribution of a preobservable $a(\omega)$ or $b(\omega)$ for 
an arbitrary context represented by a domain in the prespace $\Omega.$ Neither we can reconstruct maps 
$a(\omega)$ and $b(\omega)$. Thus the quantum theory is not a fundamental theory. It does not provide 
the complete (even statistical) description of the prespace reality. However, some statistical information
about the prespace structure can be obtained by using the quantum probabilistic formalism.

{\bf 8. On the mystery of  operator quantization.}
The origin of the operator quantization was always mysterious for me. Why the correspondence between functions and functions of operators (of the position and the momentum) provides the correct statistical description of quantum measurements? Our contextual model tells that the only reason is the coincidence of quantum averages with `real' prespace (contextual) averages for some preobservables (in particular, of the form $f(b)+ g(a)).$

Theorem 4 is only a sufficient condition for the coincidence of averages. 
But even such a result gives 
the possibility to connect the quantun Hamiltonian 
$$\hat {\cal H} =\frac{\hat a^2}{2} + V(\hat b)$$ 
with the realist preobservable ${\cal H}(a(\omega), b(\omega))=\frac{a(\omega)^2}{2} + V(b(\omega)).$ 
Quantum averages of energy expressed by the Hilbert space averages of the Hamiltonian 
$\hat {\cal H}$ coincides with averages of the realist energy preobservable ${\cal H}(\omega).$ 
However, for some contexts $C$ quantum energy observable $\hat {\cal H}$ and energy preobservable ${\cal H}$ 
have different probability distributions, see Proposition 3.

The classical space is a contextual image of the fundamental prespace $\Omega.$ 
 This is a very poor image since only a few special contexts namely space-contexts have images 
 in the classical space ${\bf R}^3$. In principle, there might be created 
 various classical spaces (corresponding to various fundamental variables)
 on the basis of the prespace $\Omega$. Human beings have been creating their own (very special) 
 classical space. Since light rays play the fundamental role in the creating of our classical 
 space it can be called {\it{electromagnetic classical space}}. So  the electromagnetic classical space
 is created on the basis on electromagnetic reduction of information. In principle there can exist
 systems which are able to perform some other reductions of information, e.g., gravitation reduction.
 They would create a {\it gravitational classical space.}

\section{Appendix on incompatible random variables}

 {\bf Proposition 4.}
 {\it Let $\{A_j\}$ and $\{ B_k\}$ be two families of subsets of some set $\Omega$ and 
 $\Omega= \cup_j A_j = \cup_k B_k$ and 
 let
 \begin{equation}
 \label{I}
 A_j B_k\ne \emptyset 
 \end{equation}
 for any pair $(j,k).$  Then
 \begin{equation}
  \label{I1}
{\mbox{Neither}}\;A_j \subset B_k \;{\mbox {nor}} \; \; B_k \subset A_j
  \end{equation}
  for any pair $(j,k).$ If $n=2$ then conditions (\ref{I}) and (\ref{I1}) are equivalent.}
  
  {\bf{Proof.}} Let (\ref{I}) hold true. Suppose that there exists $(j, k)$ such that $A_j \subset B_k.$
Thus we should have $A_j B_i= \emptyset$ for any $i\ne k.$ Let (\ref{I1}) hold true and 
let $n=2:{\cal A}=\{A_1, A_2=\Omega\setminus A_1\}$ and ${\cal B}=\{B_1, B_2=\Omega\setminus B_1\}.$ 
Suppose that, e.g., $A_1 B_1=\emptyset.$ Then we should have $A_1 \subset B_2.$ 

If $n\neq 3$ then in general the condition (\ref{I1}) does not imply the condition (\ref{I}). 
We can consider the following example. Let $\Omega=\{\omega_1, \ldots, \omega_7\}$
and let $A_1=\{\omega_1, \omega_2 \omega_3\}, A_2=\{\omega_4, \omega_5\}, 
A_3=\{\omega_6, \omega_7\}$ and $B_1=\{\omega_1, \omega_4\}, 
B_2=\{\omega_2, \omega_5, \omega_6\}, B_3=\{\omega_3, \omega_7\}.$
Here (\ref{I1}) holds true but $A_2 B_3\not= \emptyset.$

Finally, we remark that we have investigated only the case of dichotomous random variables. The general
case  is essentially more complicated from the mathematical viewpoint. In particular, not every double stochastic matrix
can be represented as the square of a unitary matrix and so on... But I think that from the 
phemenological viewpoint the case of dichotomous observables is the most important, cf., e.g., Mackey
[24] and the general quantum logic approach.

I would like to thank L. Ballentine, S. Gudder, A. Holevo, P. Lahti, B Hiley, S. Goldstein, C. Fuchs,
A. Peres, I. Volovich, R. Gill,
J. Bub, T. Maudlin,  H. Rauch, G. Emch, V. Belavkin, I. Helland
for discussions on probabilistic foundations of quantum theory.

{\bf References}

[1] A. Einstein, B. Podolsky, N. Rosen,  Phys. Rev., {\bf 47}, 777--780 (1935).

[2] J. S. Bell, {\it Speakable and unspeakable in quantum mechanics.} Cambridge Univ. Press (1987).

[3] A. Yu. Khrennikov, I.V. Volovich,   {\it Quantum Nonlocality, EPR Model, and Bell`s Theorem}, 
Proc. 3nd Sakharov Conf., Moscow, 2002, World Sci., {\bf 2,} pp.269-276 (2003).

[4] J. Bell, {\it Rev. Mod. Phys.,} {\bf 38}, 447-452 (1966).

[5] J. von Neumann, {\it Mathematical foundations
of quantum mechanics} (Princeton Univ. Press, Princeton, N.J., 1955).

[6] S. Kochen and E. Specker, {\it J. Math. Mech.}, {\bf 17}, 59-87 (1967).

[7] L. E. Ballentine, {\it Rev. Mod. Phys.}, {\bf 42}, 358--381 (1970).

[8] A. S. Wightman, Hilbert's sixth problem: mathematical treatment of the axioms of physics.
{\it Proc. Symposia in Pure Math.,} {\bf 28}, 147-233 (1976).

[9] A. S. Holevo, {\it Statistical structure of quantum theory.} Springer,
Berlin-Heidelberg (2001).

[10] L. De Broglie, {\it The current interpretation of wave mechanics,
critical study.} Elsevier Publ., Amsterdam-London-New York (1964).

[11] D. Bohm, {\it Quantum theory,} Prentice-Hall.
Englewood Cliffs, New-Jersey (1951).

[12] D.  Bohm  and B. Hiley, {\it The undivided universe:
an ontological interpretation of quantum mechanics.}
Routledge and Kegan Paul, London (1993).

[13] E. Nelson, {\it Quantum fluctuation,} Princeton Univ. Press, Princeton (1985).

[14] A. Yu. Khrennikov, Linear representations of probabilistic transformations 
induced by context transitions. {\it J. Phys.A: Math. Gen.,} {\bf 34}, 9965-9981 (2001).
quant-ph/0105059.

[15] A . Yu. Khrennikov, Quantum statistics via perturbation effects of preparation procedures. 
{\it Il Nuovo Cimento}, {\bf B 117}, N. 3, 267-281 (2002).

A. Yu Khrennikov, {\it Contextual viewpoint to quantum statistics}, hep-th/0112076.

[16] A. Yu. Khrennikov, Ensemble fluctuations and the origin of quantum probabilistic rule.
{\it J. Math. Phys.}, {\bf 43}, N. 2, 789-802 (2002).

[17] A. Yu. Khrennikov, Unification of classical and quantum probabilistic
formalisms. quant-ph/0302194.

[18] D. Hilbert, J. von Neumann, L. Nordheim, {\it Math. Ann.}, {\bf 98}, 1-30 (1927).

[19] A. Lande, {\it Foundations of quantum theory.} Yale Univ. Press (1955).

A. Lande, {\it New foundations of quantum mechanics.} Cambridge Univ. Press, Cambridge.

[20] S. P. Gudder, Special methods for a generalized probability theory.
{\it Trans. AMS,} {\bf 119}, 428-442 (1965).

S. P. Gudder, {\it Axiomatic quantum mechanics and generalized probability theory.}
Academic Press, New York (1970).

S. P. Gudder, An approach to quantum probability. Proc. Conf.
{\it Foundations of Probability and Physics,} ed. A. Khrennikov.
Quantum Prob. White Noise Anal., {\bf 13}, 147-160, WSP, Singapore (2001).

[21] L. Accardi, The probabilistic roots of the quantum mechanical paradoxes.
{\em The wave--particle dualism.  A tribute to Louis de Broglie on his 90th 
Birthday,} ed. S. Diner, D. Fargue, G. Lochak and F. Selleri
(D. Reidel Publ. Company, Dordrecht, 297--330, 1984);

L. Accardi, {\it Urne e Camaleoni: Dialogo sulla realta,
le leggi del caso e la teoria quantistica.} Il Saggiatore, Rome (1997).

[22] L. E. Ballentine, {\it Quantum mechanics} (Englewood Cliffs, 
New Jersey, 1989).

[23] L. E. Ballentine,  Interpretations of probability and quantum theory.
Proc. Conf. {\it Foundations of Probability and Physics,} ed. A. Khrennikov.
{\it Q. Prob. White Noise Anal.}, {\bf 13}, 71-84, WSP, Singapore (2001)

[24] G. W. Mackey, {\it Mathematical foundations of quantum mechanics.}
W. A. Benjamin INc, New York (1963).

[25] G. Ludwig, {\it Foundations of quantum mechanics} (Springer, 
Berlin, 1983).

[26] E. B. Davies, J. T. Lewis, {\it Comm. Math. Phys.}, {\bf 17}, 239-260 (1970).

[27] A. Yu. Khrennikov, {\it Interpretations of probability} (VSP Int. Publ., Utrecht, 1999).

[28] L. Hardy, Quantum theory from intuitively reasonable axioms. 
Proc. Conf. {\it Quantum Theory: Reconsideration
of Foundations,} ed. A. Khrennikov. 
Ser. Math. Modelling, {\bf 2}, 117-130, V\"axj\"o Univ. Press (2002).

[29] A. N. Kolmogoroff, {\it Grundbegriffe der Wahrscheinlichkeitsrech}
Springer Verlag, Berlin (1933); reprinted:
{\it Foundations of the Probability Theory}. 
Chelsea Publ. Comp., New York (1956);

[30] A. Yu. Khrennikov, On foundations of quantum theory. Proc. Int. Conf. {\it Quantum Theory: Reconsideration
of Foundations.} Ser. Math. Modelling in Phys., Engin., and Cogn. Sc., 163-196, V\"axj\"o Univ. Press,  2002.

\end{document}